\documentclass[referee]{aa} 

\usepackage{graphicx}
\usepackage{lscape}

\begin{document}

\titlerunning{Optical Photometry of Galaxies in AC\,118 Field}

   \title{Optical (VRI) Photometry in the Field of\\
 the Galaxy Cluster AC\,118 at z=0.31\thanks{Based on 
observations collected at European Southern Observatory
(ESO n. 62.O-0369, 63.O-0257, 64.O-0236)}}

    \subtitle{}

\author{G. Busarello \inst{1} 
\and
P. Merluzzi \inst{1} 
\and 
F. La Barbera \inst{2} 
\and 
M. Massarotti \inst{1}
\and 
M. Capaccioli \inst{1,2}
          }

   \offprints{G. Busarello}

   \institute{I.N.A.F., Istituto Nazionale di Astrofisica 
Osservatorio Astronomico di Capodimonte, 
Via Moiariello 16, I-80131 Napoli\\
email: gianni@na.astro.it
\and
Universit\`a Federico II, Department of Physics, Napoli
}

   \date{Received ; accepted }

   \abstract{ We present new photometric V-, R- and I-band data for
the rich galaxy cluster AC\,118 at $z=0.31$. The new photometry
covers an area of 8.6$\times$8.6 arcmin$^2$, corresponding to
2.9$\times$2.9 Mpc$^2$ (H$_0=50$ km s$^{-1}$ Mpc$^{-1}$,
q$_0=0.5$ and $\Lambda = 0$). The data have been collected
for a project aimed at studying galaxy evolution
through the color-magnitude relation and the fundamental plane.  We
provide a catalogue including all the sources ($\mathrm{N} = 1206$)
detected in the cluster field. The galaxy sample is complete to
$\mathrm{V}=22.8$ mag ($\mathrm{N_{gal}}=574$), $\mathrm{R}=22.3$ mag
($\mathrm{N_{gal}}=649$) and $\mathrm{I}=20.8$ mag
($\mathrm{N_{gal}}=419$). 
We give aperture magnitudes within a fixed aperture of 4.4
arcsec and Kron magnitudes. We also give photometric redshifts for 459
sources for which additional U- and K-band photometry is available.
We derive and discuss the V- and R-band luminosity functions. The catalogue, 
which is distributed in
electronic form, is intended as a tool for studies in galaxy
evolution.

    }

   \maketitle
%

\section{Introduction}

Rich galaxy clusters are the optimal laboratory to study the evolution
of galaxies because they can be identified over a wide range of
redshifts, thus allowing to study the properties of homogeneous
classes of galaxies as a function of cosmic time.  Rich clusters are
also the tracers of the evolution of cosmic structures and serve to set
constraints on the initial spectrum of density fluctuation.

AC\,118 (also Abell\,2744) is a rich (`Coma-like') galaxy cluster at
redshift $z=0.31$ and is one of the most widely studied clusters at
intermediate redshifts.  The first photometric (Couch \&
Newell 1984) and spectroscopic (Couch \& Sharples 1987) data
collected for this cluster exhibited a significant excess of blue
galaxies with respect to local clusters (the Butcher-Oelmer effect,
Butcher \& Oelmer 1978a, 1978b, 1984).  Couch \& Sharples (1987)
compared the distribution of the galaxy population of AC\,118 with the
morphological segregation in the Coma cluster, and suggested that the
blue galaxies in AC\,118 likely represent the progenitors of the local
S0 galaxies.

The star formation in AC\,118 was studied by Barger et
al. (\cite{BAE96}) who found that a significant fraction of galaxies
experimented secondary bursts of star formation in the last $\sim2$
Gyr.  Furthermore, since the HST image indicates that a high portion
of the star forming galaxies show signs of interaction, they 
suggested that merging phenomena of individual galaxies are responsible 
of these recent hints of star formation and that this may be consistent 
with the hierarchical merging scenario.

AC\,118 was also included in the sample of high-redshift clusters of
Stanford et al. (\cite{SED98}) who have shown that the color-magnitude
relations for AC\,118 as well as for all the other clusters is
consistent with a passive evolution of stellar populations formed at
early cosmic epochs. This result, however, is not in contrast with the 
findings of Barger et al. (\cite{BAE96}), because of the different
sample selections.
It has to be noticed, in this
respect, that samples constituted of morphologically selected
early-type galaxies (as in Stanford et al. 1998) are affected by the
`progenitor bias' (see van Dokkum \& Franx 2001), and always lead to
conclude that early-type galaxies are old.  The passive evolution
scenario is also supported by the trend with the redshift of
the K-band luminosity function, as was shown by de Propris et
al. (1999). 
Couch et al. (\cite{CBS98}) pointed out that the facts
may be much more involved than in the above schemes.  From a detailed
morphological and spectroscopic study of three clusters at z=0.31
(AC\,118, AC\,103 and AC\,114), they draw the conclusion that several
different processes drive galaxy evolution and that a significant
fraction of galaxies in a rich cluster undergo morphological changes.
The same conclusion was reached by Jones et al. (2000), who claimed
that the S0 galaxies at the redshift of AC\,118 are remnants of
early-type spirals that had star formation truncated by interaction
with the cluster environment (see also Adami et al. 1998; Poggianti et al.
1999).

AC\,118 has been the subject of several studies at different
wavelengths, from radio to X-ray (see for instance Andreani et
al. 1996; Smail et al. 1997; Allen 1998, 2000; Govoni et al. 2001).
The general picture describes this cluster as an assembling of
different systems, with a dynamical activity suggested by a peculiar
velocity distribution with two different peaks (Girardi \& Mezzetti
2001), and by the strong discrepancy between the masses derived from
X-ray observations and from gravitational lensing (Allen
1998). Finally, AC\,118 will be one main target in the first
operations of the Advanced Camera for Surveys of HST.
  
This paper presents new photometric data for AC\,118 based on ESO-NTT
imaging in the R, V and I wavebands. These data were collected in the
framework of a project aimed at studying the structural properties of
the cluster galaxies (Busarello et al. 2002; La Barbera et al. 2002), 
the color-magnitude relation (Merluzzi et al. 2002)
and at deriving the fundamental plane. In this paper we also
determine the photometric redshifts and briefly discuss the V and R
luminosity functions.  The paper is organized as follows. Sect. 2
includes a description of the data and of the data reduction and
photometric calibration. In Sect. 3 we describe the aperture
photometry. Sect. 4 deals with the
determination of the photometric redshifts. The photometric catalogue
is presented in Sect. 5, while the catalogue itself is available in
electronic form only. In Sect. 6 we shortly discuss the V and R
luminosity functions, and in Sect. 7 we give a summary of the
paper. In this work we assume H$_0=50$ km s$^{-1}$ Mpc$^{-1}$,
q$_0=0.5$ and $\Lambda = 0$.


\section{Observations, Data Reduction and Photometric Calibration}

The observations were carried out at the ESO New Technology Telescope
with the EMMI instrument during three runs (October 1998, October 1999 and 
September 2000). A field of $8.6^\prime \times 8.6^\prime$ (2.9$\times$2.9 Mpc) 
was observed in V-, R- and I-bands towards the galaxy cluster AC\,118. 
The total integration time, the pixel scale, and the average seeing for each 
band are given in Table~\ref{dataset}.

\begin{table}
\caption[]{Relevant information on the observations.}
\label{dataset}
$$
\begin{array}{cccc}
\hline
\noalign{\smallskip}
\mathrm{Band} & \mathrm{T_{exp}} & \mathrm{Scale} & \mathrm{Seeing} \\
&   \mathrm{ksec} & ''/\mathrm{pxl} & '' \\
\noalign{\smallskip}
\hline
\noalign{\smallskip}
{\rm V} & 2.7 & 0.27 & 1.2 \\
{\rm R} & 1.8 & 0.27 & 1.0 \\
{\rm I} & 1.2 & 0.27 & 1.2  \\
\noalign{\smallskip}
\hline
\end{array}
$$
\end{table}

The data reduction was performed with the MIDAS and IRAF packages. 
Bias and flat-field corrections were achieved by the standard procedures. 
For the V- and R-bands the flat-field was obtained by combining twilight 
and dome flats, while for the I-band only the dome flat was used. 
In order to add the different exposures and to reject cosmic rays 
we used the IRAF task IMCOMBINE with the CRREJECT algorithm.
The final images presented a uniform background with fluctuations smaller than
2\% of the mean value.

The photometric calibration was performed into the
Johnson--Kron--Cousins photometric system.  Since the sky conditions
were not photometric during the observations, we could not use
comparison standard fields, but we proceeded as follows.

\begin{figure*}
\centering
\includegraphics[angle=0,width=14cm,height=14cm]{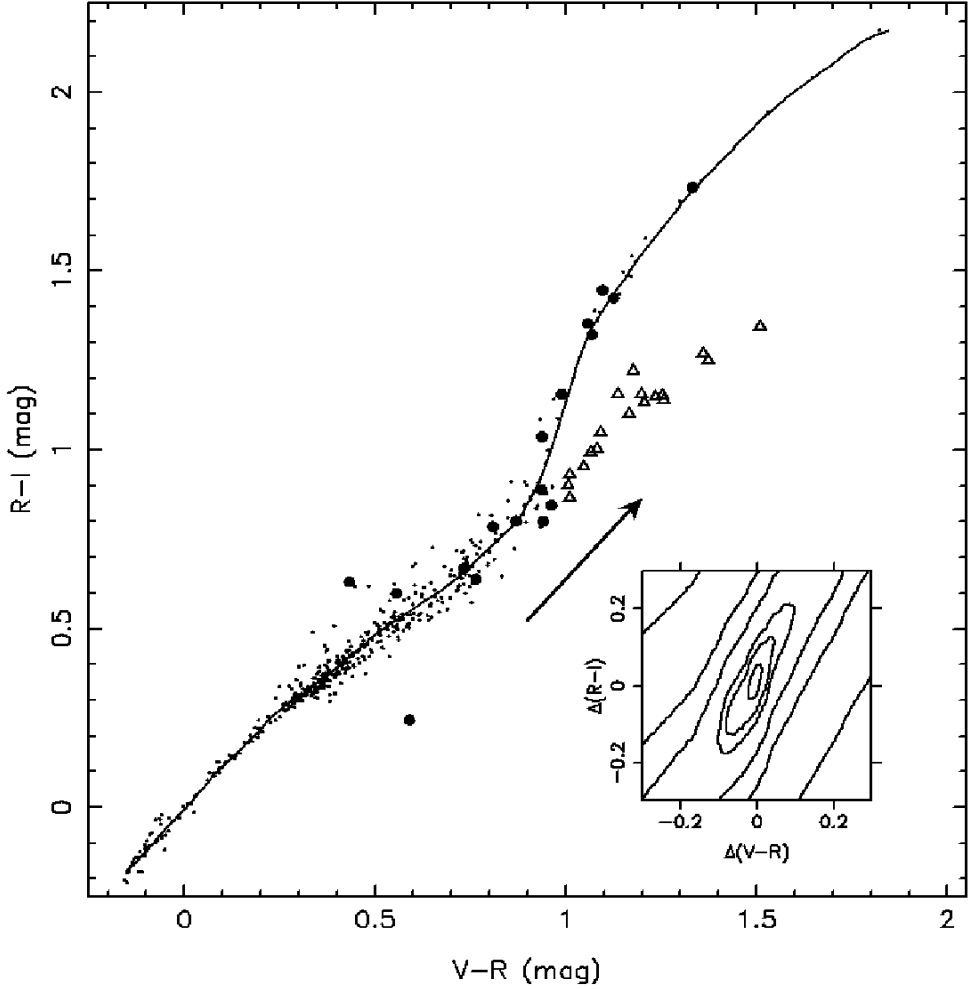}
\caption{Calibration of the V- and I-band images by means of the
Landolt stars.  The Landolt stars used to obtain the smooth curve are
represented as dots in the (V-R, R-I) diagram. The triangles are the
Landolt stars with strong galactic absorption. The reddening direction
(Schlegel et al. 1998) for these points is indicated by an arrow.
Stars in the NTT images after the calibration are plotted as filled
circles.  Notice that some points show a larger deviation from the
locus of the Landolt stars due to their larger photometric errors.
The contour plot of the $\chi^2$ of the fit is shown in the lower
right of the figure. Notice that the scales of the bigger and the
smaller panels are the same.  }
\label{COLFIT}
\end{figure*}

For the R-band, we used the aperture photometry for $\mathrm{N}=92$
sources in the cluster area provided to us by A. Stanford (see
Stanford et al. \cite{SED98}, hereafter SED98).  Instrumental
magnitudes within an aperture of the same diameter ($5''$ ) as in SED98
were derived from our image by using the SExtractor software (Bertin
$\&$ Arnouts 1996) and were compared with the magnitudes of SED98.  Since
the FWHM of the SED98 images was $\sim 1 ''$, no seeing correction was
applied for the matching with our data.  The accuracy on the zero
point was derived by adding in quadrature the uncertainty on the
offset between the two sets of magnitudes with the uncertainty on  the
photometric calibration of SED98. The total uncertainty on the zero
point amounts to 0.03 mag (1$\sigma$ standard interval).

For the V- and I-bands, the photometric calibration was performed by
matching the colors of $\mathrm{N}=17$ stars in the NTT images with
the colors of the stars from the catalogue of Landolt (1992). We
proceeded as illustrated in Fig.~\ref{COLFIT}. The distribution of the
Landolt stars in the (V-R, R-I) diagram was described by a smooth
curve.  Some stars were not considered since they do not lie in the
same locus of the other Landolt stars due to their high galactic
extinction ($\mathrm{E(B-V)} > 0.4$, see Schlegel et al. 1998).  We
verified that our curve agrees with the curve of Caldwell et
al. (1993). The $v-R$ and $R-i$ colors of the stars in the NTT images,
where $v$ and $i$ are the instrumental magnitudes, were fitted to the
smooth curve and the computed shifts were added to the zero point of the 
R-band image.

The accuracies on the zero points were estimated by taking into
account 1) the presence of a possible color term in the calibration,
2) the uncertainty on the fitting solution and 3) the uncertainty
on the R-band photometric calibration. To estimate the first
source of uncertainty, we subtracted to the colors of the Landolt
stars the color term in the equations of the EMMI Calibration Plan
\footnote{See http://www.ls.eso.org/lasilla/Telescopes/NEWNTT/EMMI/}
and estimated the change of the fitting solution. The second source of
uncertainty was estimated by the bootstrap technique taking into
account the uncertainties on the instrumental magnitudes.  The
uncertainties on the colors of the Landolt stars were not considered,
since their effect on the computation of the smooth curve turned out
to be negligible. The uncertainties on the zero points are summarized
in Table~\ref{paramobs}. Since the uncertainty on the R-band zero point
turned out to be the major source of error, we neglected the correlation
between $\mathrm{\Delta ZP_V}$ and $\mathrm{\Delta ZP_I}$ introduced by
the fitting procedure.

\begin{table}
\caption[]{Parameters of the photometry. For each waveband
$\Delta_{\mathrm{ZP}}$,
$\mathrm{M_C}$ and $\mathrm{N_g}$ are the uncertainty on the zero
point, the completeness magnitude, and the total number of galaxies 
with aperture magnitude brighter than $\mathrm{M_C}$. 
We identify as galaxies the objects with stellar index value smaller 
than 0.9.}
\label{paramobs}
$$
\begin{array}{cccc}
\hline
\noalign{\smallskip}
$$\mathrm{Band}$$ & $$\Delta_{\mathrm{ZP}}$$ & $$\mathrm{M_C}$$ & 
$$\mathrm{N_g}$$ \\
& $$\mathrm{mag}$$ & $$\mathrm{mag}$$ & \\
\noalign{\smallskip}
\hline
\noalign{\smallskip}
$$\mathrm{V}$$ & 0.04 & 22.8 & 574\\
$$\mathrm{R}$$ & 0.03 & 22.3 & 649\\
$$\mathrm{I}$$ & 0.05 & 20.8 & 419\\
\noalign{\smallskip}
\hline
\end{array}
$$
\end{table}

\begin{figure*}
\centering
\includegraphics[angle=0,width=\textwidth,height=\textwidth]{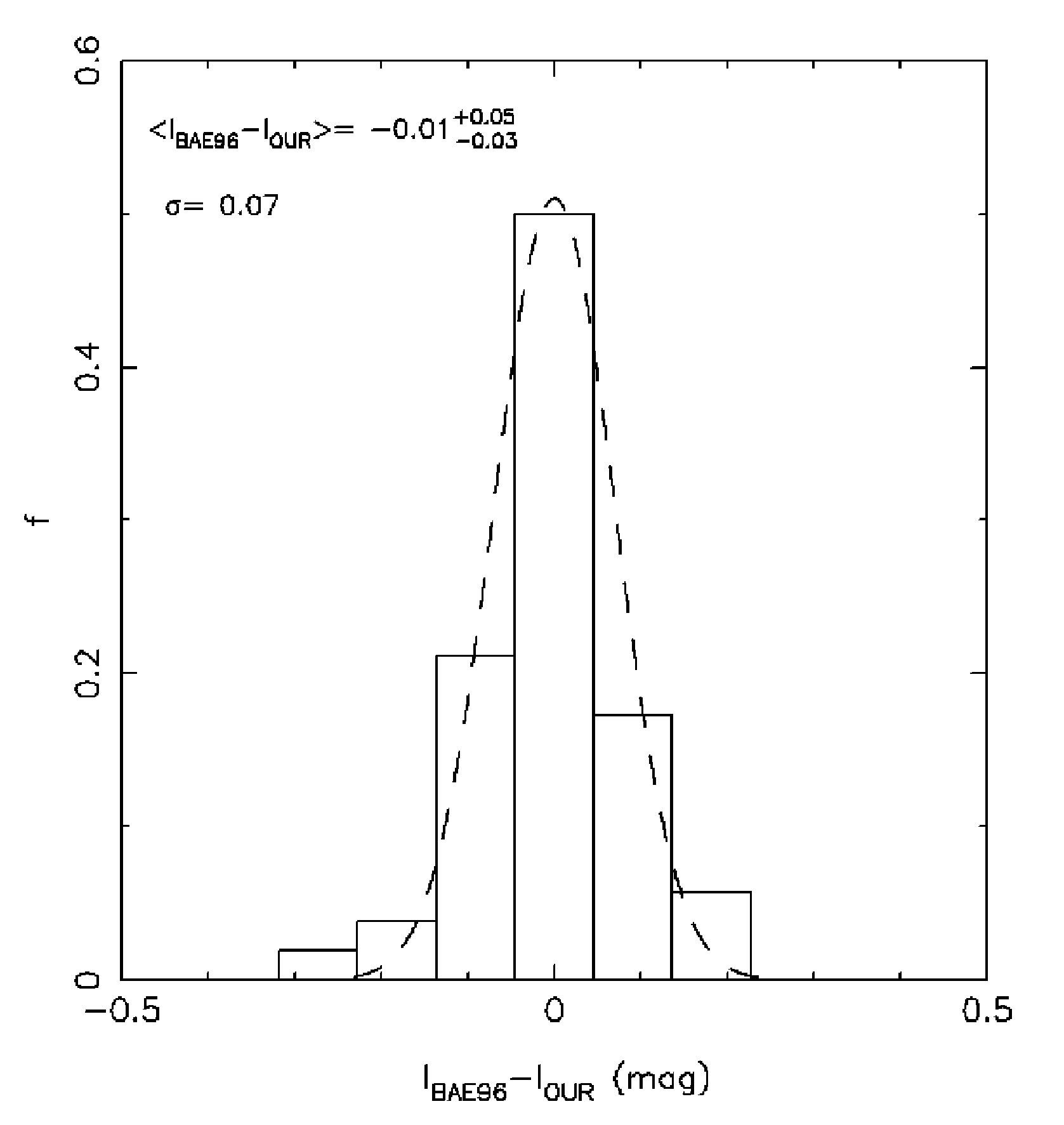}
\caption{Comparison of our I-band photometry with BAE96: distribution of the 
differences between BAE96 and our magnitudes.
}
\label{BAE}
\end{figure*}

To check the method used for the photometric calibration, we compared our 
photometry with the magnitudes in the I-band of Barger et al. \cite{BAE96} 
(hereafter BAE96). The comparison is shown in Fig.~\ref{BAE}. 
The difference between the two sets of magnitudes
is fully consistent with zero within the estimated uncertainty on the
I-band zero point. The dispersion amounts to 0.07 mag, in agreement 
with the various sources of uncertainty.

\section{Aperture Photometry}

A photometric catalogue was derived for each image  by the software SExtractor 
(Bertin
$\&$ Arnouts \cite{BER96}).  For each object we measured the magnitude
within a fixed aperture of diameter $4.4 ''$, corresponding to $\sim$25 kpc
at $z=0.31$, and the Kron magnitude (Kron 1980), for which an
adaptive aperture is used with diameter equal to $\alpha \cdot r_c$,
where $r_c$ is the Kron radius and $\alpha$ is a constant. Following
Bertin $\&$ Arnouts (\cite{BER96}), we chose $\alpha=2.5$, for which it
is expected that the Kron magnitude encloses $\sim 94 \%$ of the total
flux of the source. The uncertainties on the magnitudes were obtained
by adding in quadrature the uncertainties estimated by SExtractor with
the uncertainties on the zero points.

\begin{figure*}
\centering
\includegraphics[angle=0,width=\textwidth]{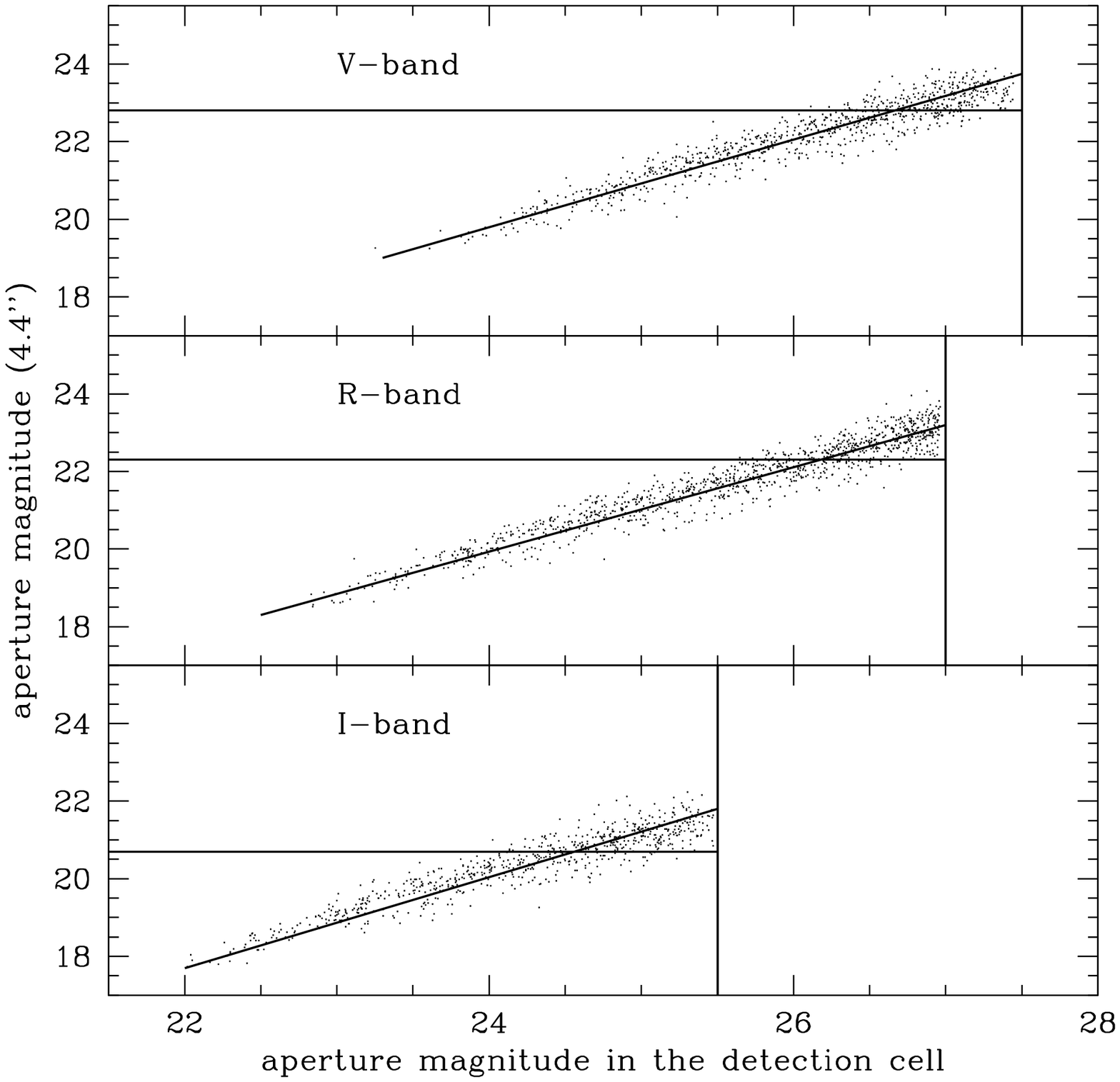}
\caption{Completeness limits of the photometric catalogues.  The
magnitudes within the fixed aperture are plotted against the
magnitudes in the detection aperture for the V-, R- and I-band
catalogues (upper, middle and lower panels). The vertical and
horizontal lines correspond to the detection threshold and the
completeness magnitude respectively.  }
\label{compl}
\end{figure*}

The completeness of the optical catalogues was estimated following the
method of Garilli et al. (\cite{GAR99}) that consists in the
determination of the magnitude at which the objects start to be lost
since they are below the brightness threshold in the detection cell.
To estimate the completeness limit, the magnitudes in the detection cell
were computed and compared to the magnitudes in the fixed aperture.
The comparisons are shown in Fig.~\ref{compl}, where the vertical
lines correspond to the detection threshold and the horizontal lines
mark the completeness limit which takes into account the scatter of the 
relation. The completeness magnitudes are reported in Table~\ref{paramobs}.

\section{The photometric redshifts}

The optical V-, R- and I-band photometry, together with available data
in the U- and K-band was used to determine the cluster membership of
the sources in the photometric catalogue by the photometric redshift
technique.  AC\,118 was observed in the U-band by BAE96 at
the ESO-NTT on August 1993. The images were retrieved from the ESO
NTT Science Archive and reduced following the standard procedures. For the
photometric calibration we used the aperture magnitudes given in BAE96
(see their Table~4). The K-band observations have been performed at
ESO-NTT with the SOFI instrument.  The K-band photometry
consists of a total integration of $\sim$4 hours over a single SOFI pointing
of $5^\prime \times 5^\prime$ which, because of the jittering, covers a 
field of $ 6^\prime \times 6^\prime$ (the reduction of the SOFI image is
described in Andreon 2001). The UVRIK catalogues were
cross-correlated, resulting in a final list of $\mathrm{N} = 1206$
sources detected in at least one optical band. The photometric
redshifts were derived for the $\mathrm{N} = 459$ objects with 
photometry available in all the wavebands.

We estimated the  photometric redshifts according to the Spectral Energy
Distribution fitting method (see Massarotti et al. 2001a, b, and
references therein). Redshifts are obtained by comparing observed
galaxy fluxes, $f^\mathrm{obs}_i$ at the $i$--th photometric band,
with a library of reference fluxes, $f^\mathrm{templ}_i(z, T)$,
depending on redshift $z$ and on a suitable set of parameters $T$, that
account for galaxy morphological type, age, metallicity, etc. 

We looked for redshifts in the range $z \in [0.0, 1.0]$ with a step of
0.01, and introduced a further constraint in the fit by imposing that the
age of the best template cannot be older than the age of the universe
at the selected redshift according to the adopted cosmology.
Model galaxy spectra are provided by the code of Bruzual \& Charlot
(\cite{BRC93}). The adopted templates consist of models with 
a Scalo (1986) IMF and with SFR histories of the form $SFR(t) \propto 
e^{-t/\tau}$ to match different Hubble type colors as a function of 
time. Choosing $\tau = 1, 4, 15$ Gyr, we obtain at $t=12.0$ Gyr
colors of $E/S0$, $Sa/Sb$, and $Sc/Sd$ that match those
of nearby galaxies. Template spectra evolution is
followed in the time interval $t \in [1.0,12.0]$ Gyr.
To allow for different metallicities of early-type galaxies we introduced
templates  for $E/S0$ with $Z/Z_\odot=$0.2, 0.4, 1 and 2.5.

\begin{figure*}
\centering
\includegraphics[angle=0,width=\textwidth,height=13cm]{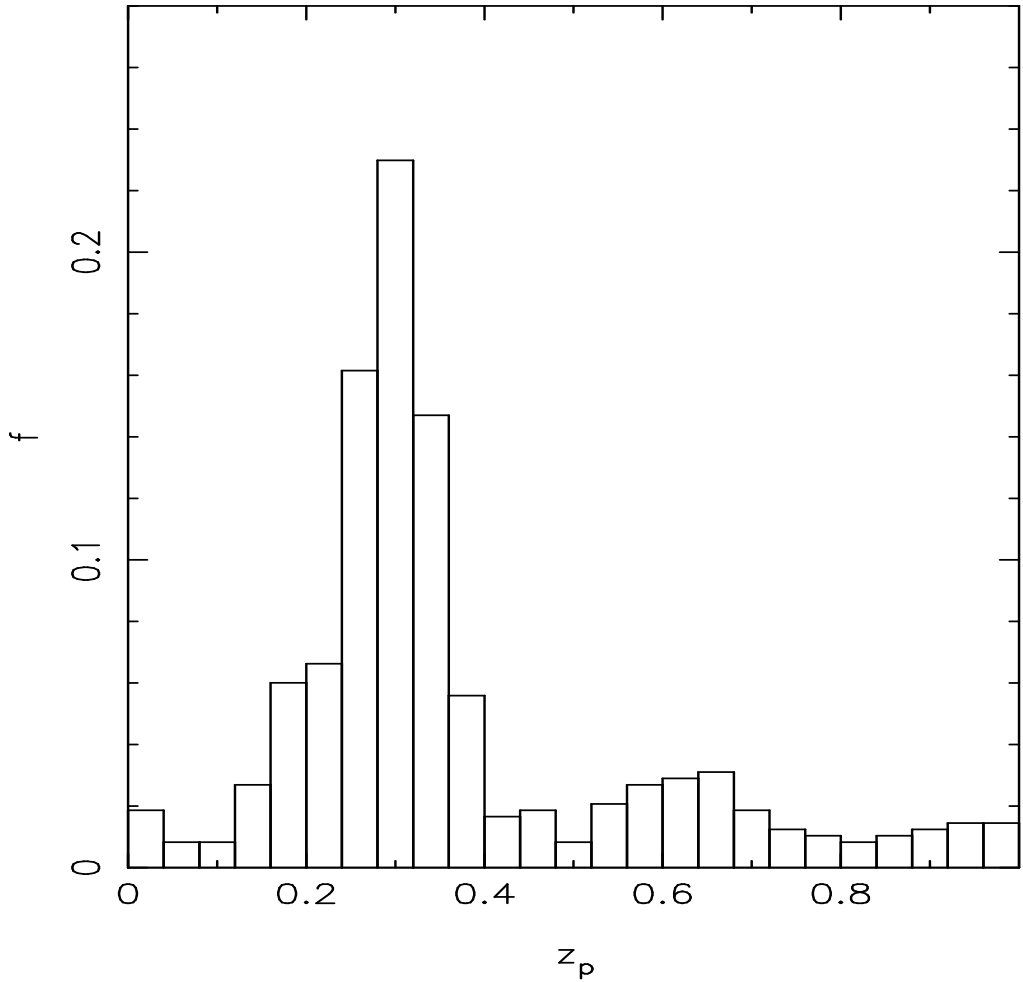}
\caption{Distribution of the photometric redshifts $\mathrm{z_p}$ for 
the N=459 sources with available UVRIK photometry.
}
\label{dis}
\end{figure*}

The distribution of photometric redshifts is shown in Fig.~\ref{dis}.
The distribution is dominated by the peak around the cluster redshift 
$z \sim 0.3$, indicating that most of the galaxies in the $6^\prime 
\times 6^\prime$ field are actually members of AC\,118. 
The FWHM of the peak $\Delta z \sim 0.1$ is determined by the intrinsic
redshift distribution of the cluster members and on the errors on
photometric redshifts. 

\begin{figure*}
\centering
\includegraphics[angle=0,width=\textwidth,height=13cm]{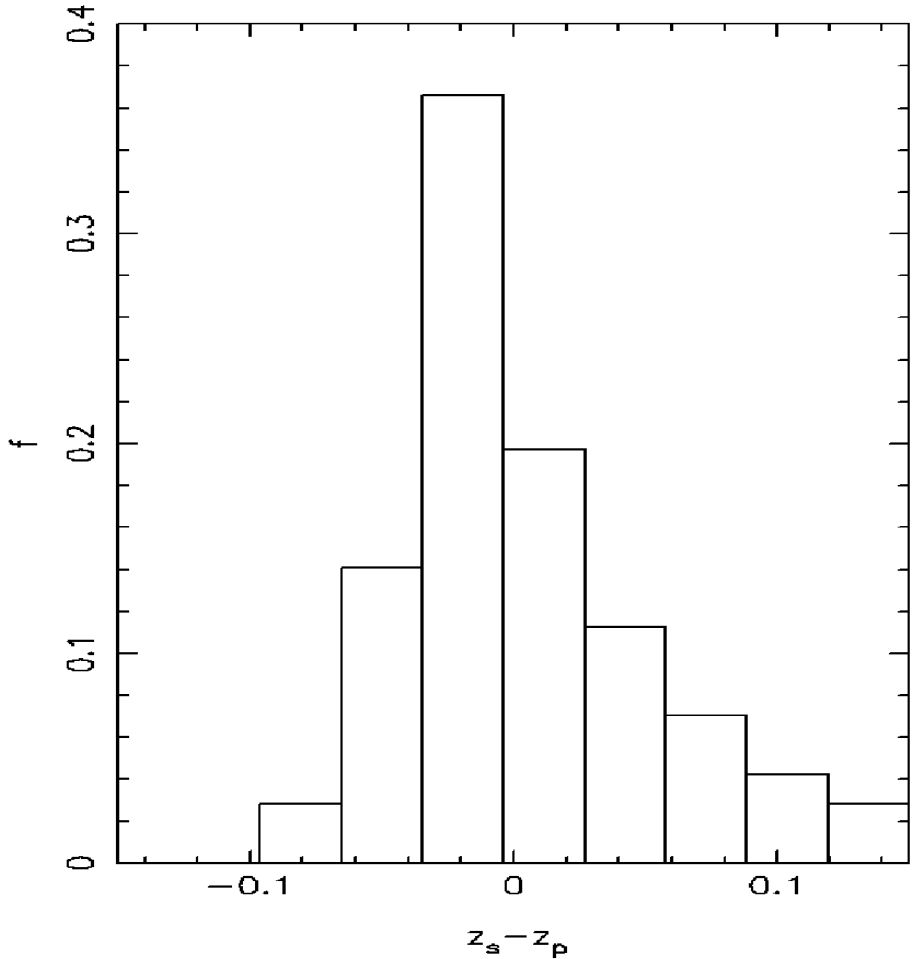}
\caption{Comparison of spectroscopic $\mathrm{z_s}$ and photometric 
redshifts $\mathrm{z_p}$ for the N=75 galaxies with available spectroscopy.
}
\label{photspec}
\end{figure*}

To gain insight on the precision of the redshift estimates, we compare
in Fig.~\ref{photspec} photometric ($z_\mathrm{p}$) and spectroscopic
($z_\mathrm{s}$) redshifts for galaxies in a bright sub--sample of the
cross-correlated catalogue (see Couch \& Sharples 1987; Busarello et al. 
2002, in preparation).  We
found that the mean difference of photometric and spectroscopic
redshifts is consistent with zero and that the dispersion amounts to
0.04.  We notice that $95 \%$ of the spectroscopically confirmed cluster 
members  are found in the range $z_\mathrm{p} \in [0.24,0.38]$.
On the other hand, since galaxies in the spectroscopic sample
represent the bright tail of the cross-correlated catalogue, the
comparison of $z_\mathrm{p}$ and $z_\mathrm{s}$ does not inform on the
amplitude of the error due to photometric
uncertainties for the intermediate and faint galaxy population. In order to
analyze the role of measurement errors on the redshift estimates, we
performed numerical simulations to generate N=500 copies of the
cross-correlated catalogue, obtained by adding to each galaxy flux a
number randomly extracted from the gaussian function of the
photometric uncertainty In this way, we associated to each galaxy a
distribution of N=500 simulated redshifts ($z_\mathrm{sim}$) and
calculated the 68\% percentile interval ($\Delta z_{68}$) around the
median value of the distribution.
Finally we defined a galaxy as a cluster member if the interval $\Delta
z_{68}$ overlaps the range $z \in [0.24,0.38]$. In the
cross-correlated catalogue N= 329 galaxies (i.e. 72\% of the total) 
turned out to satisfy this selection criterion.

\section{Description of the Catalogue}

The photometric quantities relative to the  $\mathrm{N}=1206$ sources in the 
V-, R- and I-band images are included in the catalogue, a portion of which
is shown in Table~\ref{catalog}\footnote{
The whole catalogue will be available in electronic form at the Centre de
Donn\'ees astronomiques de Strasbourg: http://cdsweb.u-strasbg.fr/ 
}. The sample consists in all the objects
detected in at least band, that are brighter than the detection limit.
When an object is not detected, the corresponding
columns in the table are filled with dots. \ 
The table is organized as follows. \\
Column 1: running number in the present catalogue. \\
Columns 2, 3: right ascension and declination (2000). 
The astrometric solution was computed by using a list of stars
from the USNO catalogue. The root mean square of the residuals to
the astrometric solution amounts to 0.18 arcsec. \\
Columns 4, 5: V-band magnitude within the aperture of 
diameter $4.4 ''$ and corresponding uncertainty.  \\
Columns 6, 7: the same quantities as in columns 4, 5 for the R-band. \\
Columns 8, 9: the same quantities as in columns 4, 5 for the I-band. \\
Columns 10, 11: Kron magnitude in the V-band and corresponding 
uncertainty. \\
Columns 12, 13: the same quantities as in columns 10, 11 for the R-band. \\
Columns 14, 15: the same quantities as in columns 10, 11 for the I-band. \\
Columns 16, 17, 18: Kron radius in arcsec for the V-, R- and I-band 
respectively. When the Kron radius is smaller than 3.5 pixels, an aperture
with this diameter, that correspond to $\sim 1''$, is adopted. \\
Column 19: stellar index. When available for more than one image, we report 
the value relative to the image with the best seeing. \\
Column 20: photometric redshift. The upper and lower values correspond
 to percentile intervals of $68\%$ (see Sect.~4.3). 
  
\begin{table}
\caption[]{Catalogue of the sources in the V-, R- and I-band images.}
\label{catalog}
\scriptsize
\tiny
$$
\begin{array}{cccccccccccccccccccc}
\hline
\noalign{\smallskip}
  \mathrm{ID. \#}   &   \mathrm{R.A. (J2000)}   &   \mathrm{DEC. (J2000)}   &   \mathrm{V_a}   &   \mathrm{\delta V_a}   &   \mathrm{R_a}   &   \mathrm{\delta R_a}   &   \mathrm{I_a}   &   \mathrm{\delta I_a}   &   \mathrm{V_c}   &   \mathrm{\delta V_c}   &   \mathrm{R_c}   &   \mathrm{\delta R_c}   &   \mathrm{I_c}   &   \mathrm{\delta I_c}   &   \mathrm{r_c^{V}}   &   \mathrm{r_c^{R}}   &   \mathrm{r_c^{I}}   &   \mathrm{S/G}   &   \mathrm{z_f}     \\
  300 & 0\!:\!14\!:\!12.201 & -30\!:\!24\!:\!44.61 &   23.264 &   0.113 &   22.550 &   0.060 &    ... &   ... &   23.306 &   0.086 &   22.555 &   0.055 &    ... &   ... &   0.988 &   1.053 &   ... &   0.94 &   ...   \\
  301 & 0\!:\!14\!:\!07.669 & -30\!:\!24\!:\!45.46 &    ... &   ... &   22.846 &   0.075 &   21.324 &   0.103 &    ... &   ... &   22.839 &   0.074 &   21.318 &   0.090 &   ... &   1.353 &   1.102 &   0.89 &   ...   \\
  302 & 0\!:\!14\!:\!34.310 & -30\!:\!24\!:\!37.97 &   21.797 &   0.049 &   20.981 &   0.032 &   20.164 &   0.059 &   21.769 &   0.051 &   20.918 &   0.034 &   20.150 &   0.061 &   1.069 &   1.050 &   1.067 &   0.17 &   ...   \\
  303 & 0\!:\!14\!:\!25.755 & -30\!:\!24\!:\!40.96 &   22.831 &   0.081 &   21.939 &   0.042 &   21.117 &   0.090 &   22.879 &   0.068 &   21.941 &   0.041 &   21.122 &   0.086 &   0.945 &   0.986 &   1.210 &   0.93 &   0.28^{0.65}_{0.28}  \\
  304 & 0\!:\!14\!:\!21.318 & -30\!:\!24\!:\!38.51 &   22.081 &   0.054 &   21.201 &   0.034 &   20.482 &   0.065 &   22.058 &   0.059 &   21.177 &   0.034 &   20.482 &   0.065 &   1.288 &   0.945 &   1.075 &   0.32 &   0.30^{0.31}_{0.29}  \\
  305 & 0\!:\!14\!:\!39.343 & -30\!:\!24\!:\!41.01 &   22.928 &   0.087 &   22.394 &   0.054 &    ... &   ... &   22.878 &   0.095 &   22.397 &   0.066 &    ... &   ... &   1.369 &   1.247 &   ... &   0.01 &   ...   \\
  306 & 0\!:\!14\!:\!23.016 & -30\!:\!24\!:\!22.44 &   21.651 &   0.047 &   20.720 &   0.032 &   19.981 &   0.056 &   21.317 &   0.051 &   19.704 &   0.034 &   19.782 &   0.060 &   1.574 &   1.550 &   1.445 &   0.09 &   0.31^{0.32}_{0.30}  \\
  307 & 0\!:\!14\!:\!17.751 & -30\!:\!24\!:\!31.76 &   20.990 &   0.042 &   20.139 &   0.031 &   19.358 &   0.052 &   20.810 &   0.043 &   20.000 &   0.031 &   19.264 &   0.053 &   1.104 &   0.956 &   0.972 &   0.05 &   0.32^{0.33}_{0.31}  \\
  308 & 0\!:\!14\!:\!23.638 & -30\!:\!24\!:\!30.17 &   20.343 &   0.041 &   19.916 &   0.030 &   19.333 &   0.052 &   20.200 &   0.041 &   19.729 &   0.031 &   19.200 &   0.053 &   1.013 &   1.050 &   0.945 &   0.03 &   0.03^{0.15}_{0.01}  \\
  309 & 0\!:\!14\!:\!15.290 & -30\!:\!24\!:\!39.78 &   23.005 &   0.092 &   22.198 &   0.048 &   21.240 &   0.097 &   22.996 &   0.090 &   22.205 &   0.045 &   21.118 &   0.105 &   1.261 &   0.945 &   1.669 &   0.80 &   0.65^{0.71}_{0.55}  \\
  310 & 0\!:\!14\!:\!08.704 & -30\!:\!24\!:\!41.51 &    ... &   ... &   23.045 &   0.088 &    ... &   ... &    ... &   ... &   23.077 &   0.090 &    ... &   ... &   ... &   1.382 &   ... &   0.83 &   ...   \\
  311 & 0\!:\!14\!:\!37.739 & -30\!:\!24\!:\!33.47 &   22.862 &   0.083 &   21.764 &   0.039 &   20.816 &   0.075 &   22.596 &   0.095 &   21.573 &   0.049 &   20.716 &   0.085 &   1.823 &   1.393 &   1.463 &   0.01 &   ...   \\
  312 & 0\!:\!14\!:\!21.395 & -30\!:\!24\!:\!34.42 &   21.793 &   0.049 &   20.987 &   0.032 &   20.330 &   0.062 &   21.761 &   0.050 &   20.957 &   0.033 &   20.352 &   0.060 &   1.048 &   0.945 &   0.945 &   0.77 &   0.25^{0.26}_{0.24}  \\
  313 & 0\!:\!14\!:\!27.700 & -30\!:\!24\!:\!37.68 &   22.433 &   0.063 &   22.110 &   0.046 &    ... &   ... &   22.456 &   0.065 &   22.098 &   0.046 &    ... &   ... &   1.177 &   1.107 &   ... &   0.72 &   ...   \\
  314 & 0\!:\!14\!:\!18.265 & -30\!:\!24\!:\!39.77 &    ... &   ... &   22.816 &   0.073 &   21.531 &   0.120 &    ... &   ... &   22.861 &   0.071 &   21.517 &   0.081 &   ... &   1.280 &   0.945 &   0.93 &   ...   \\
  315 & 0\!:\!14\!:\!19.807 & -30\!:\!24\!:\!35.89 &   22.520 &   0.067 &   21.538 &   0.036 &   20.580 &   0.068 &   22.526 &   0.064 &   21.521 &   0.038 &   20.486 &   0.071 &   1.067 &   1.088 &   1.353 &   0.93 &   0.62^{0.67}_{0.47}  \\
  316 & 0\!:\!14\!:\!21.669 & -30\!:\!24\!:\!01.49 &   21.100 &   0.043 &   20.120 &   0.031 &   19.348 &   0.052 &   20.645 &   0.046 &   19.585 &   0.032 &   19.059 &   0.055 &   1.201 &   1.010 &   1.226 &   0.03 &   0.67^{0.68}_{0.66}  \\
  317 & 0\!:\!14\!:\!21.535 & -30\!:\!24\!:\!36.62 &   23.251 &   0.112 &   22.438 &   0.056 &    ... &   ... &   23.185 &   0.119 &   22.465 &   0.060 &    ... &   ... &   1.563 &   1.169 &   ... &   0.40 &   ...   \\
  318 & 0\!:\!14\!:\!07.718 & -30\!:\!24\!:\!28.74 &   20.801 &   0.041 &   20.043 &   0.031 &   19.321 &   0.052 &   20.478 &   0.043 &   19.773 &   0.031 &   18.982 &   0.054 &   1.247 &   1.023 &   1.318 &   0.03 &   0.23^{0.25}_{0.19}  \\
  319 & 0\!:\!14\!:\!22.839 & -30\!:\!24\!:\!33.73 &   22.781 &   0.079 &   21.784 &   0.040 &   20.782 &   0.074 &   22.654 &   0.085 &   21.437 &   0.045 &   20.487 &   0.079 &   1.628 &   1.574 &   1.779 &   0.83 &   0.66^{0.70}_{0.56}  \\
  320 & 0\!:\!14\!:\!25.968 & -30\!:\!24\!:\!35.48 &   21.924 &   0.051 &   21.363 &   0.035 &   21.157 &   0.092 &   21.946 &   0.048 &   21.375 &   0.034 &   21.081 &   0.074 &   0.945 &   0.972 &   0.945 &   0.98 &   0.38^{0.57}_{0.33}  \\
  321 & 0\!:\!14\!:\!17.453 & -30\!:\!24\!:\!34.37 &   21.572 &   0.046 &   21.155 &   0.033 &   20.587 &   0.068 &   21.565 &   0.046 &   21.154 &   0.034 &   20.466 &   0.071 &   1.099 &   0.964 &   1.488 &   0.98 &   ...   \\
  322 & 0\!:\!14\!:\!27.626 & -30\!:\!24\!:\!30.31 &   21.980 &   0.052 &   21.148 &   0.033 &   20.487 &   0.065 &   21.842 &   0.060 &   20.819 &   0.037 &   20.407 &   0.071 &   1.342 &   1.434 &   1.320 &   0.04 &   0.25^{0.27}_{0.25}  \\
  323 & 0\!:\!14\!:\!37.415 & -30\!:\!24\!:\!35.82 &    ... &   ... &   23.362 &   0.115 &    ... &   ... &    ... &   ... &   23.302 &   0.130 &    ... &   ... &   ... &   1.820 &   ... &   0.64 &   ...   \\
  324 & 0\!:\!14\!:\!25.408 & -30\!:\!24\!:\!34.93 &   22.993 &   0.091 &   22.274 &   0.051 &   21.542 &   0.121 &   22.992 &   0.086 &   22.290 &   0.047 &   21.545 &   0.095 &   1.188 &   0.945 &   1.013 &   0.28 &   0.19^{0.63}_{0.18}  \\
  325 & 0\!:\!14\!:\!22.013 & -30\!:\!24\!:\!27.98 &   21.819 &   0.049 &   20.892 &   0.032 &   19.995 &   0.057 &   21.341 &   0.054 &   20.519 &   0.034 &   19.663 &   0.066 &   1.588 &   1.021 &   1.391 &   0.77 &   0.32^{0.66}_{0.32}  \\
  326 & 0\!:\!14\!:\!20.126 & -30\!:\!24\!:\!34.18 &    ... &   ... &   22.716 &   0.068 &    ... &   ... &    ... &   ... &   22.725 &   0.069 &    ... &   ... &   ... &   1.164 &   ... &   0.73 &   ...   \\
  327 & 0\!:\!14\!:\!21.671 & -30\!:\!24\!:\!26.52 &   22.873 &   0.084 &   21.822 &   0.040 &   20.875 &   0.078 &   22.405 &   0.080 &   21.145 &   0.054 &   20.280 &   0.074 &   1.898 &   1.291 &   2.049 &   0.35 &   0.53^{0.65}_{0.42}  \\
  328 & 0\!:\!14\!:\!38.466 & -30\!:\!24\!:\!18.22 &   20.549 &   0.041 &   19.899 &   0.030 &   19.209 &   0.052 &   20.429 &   0.042 &   19.717 &   0.031 &   19.075 &   0.053 &   1.069 &   1.080 &   1.053 &   0.03 &   ...   \\
  329 & 0\!:\!14\!:\!15.370 & -30\!:\!24\!:\!22.53 &   21.629 &   0.046 &   20.788 &   0.032 &   19.951 &   0.056 &   21.351 &   0.054 &   20.511 &   0.034 &   19.805 &   0.059 &   1.272 &   1.064 &   0.961 &   0.02 &   0.92^{1.00}_{0.77}  \\
  330 & 0\!:\!14\!:\!11.546 & -30\!:\!24\!:\!30.51 &   22.121 &   0.054 &   21.205 &   0.034 &   20.516 &   0.066 &   22.112 &   0.057 &   21.159 &   0.035 &   20.523 &   0.063 &   1.161 &   1.069 &   0.945 &   0.46 &   0.28^{0.30}_{0.27}  \\
\noalign{\smallskip}
\hline
\end{array}
$$
\normalsize 
\end{table}

\section{R- and V-band Luminosity Functions}

To obtain the cluster R and V luminosity functions (LFs) we did not
make use of the cluster membership information derived from the
photometric redshifts, but instead we statistically subtracted the
background contribution in each band. In this way all the photometric
data up to the magnitude limit can be used in each band.  We did not
derive the LF in the I-band because the relative data are not deep
enough to estimate the slope of the LF faint end.  The background
counts were provided to us by S. Arnouts in R- and V-band from the
ESO--Sculptor Survey (Arnouts et al. 1997). This survey,
1.53$^\circ$(R.A.)$\times$0.24$^\circ$(Dec) wide, was carried out in a
region of sky close to the AC\,118 field and with the same instrument.
The reduction of those data was performed by using the same
software packages and following an analogue procedure as in the
present work. In particular, in both works the photometric
catalogues were obtained by means of the SExtractor software and the
total magnitudes, considered for the galaxy counts, were estimated
with the Kron magnitude defined by adopting the same adaptive aperture.

\begin{figure}
\resizebox{\hsize}{18cm}{\includegraphics{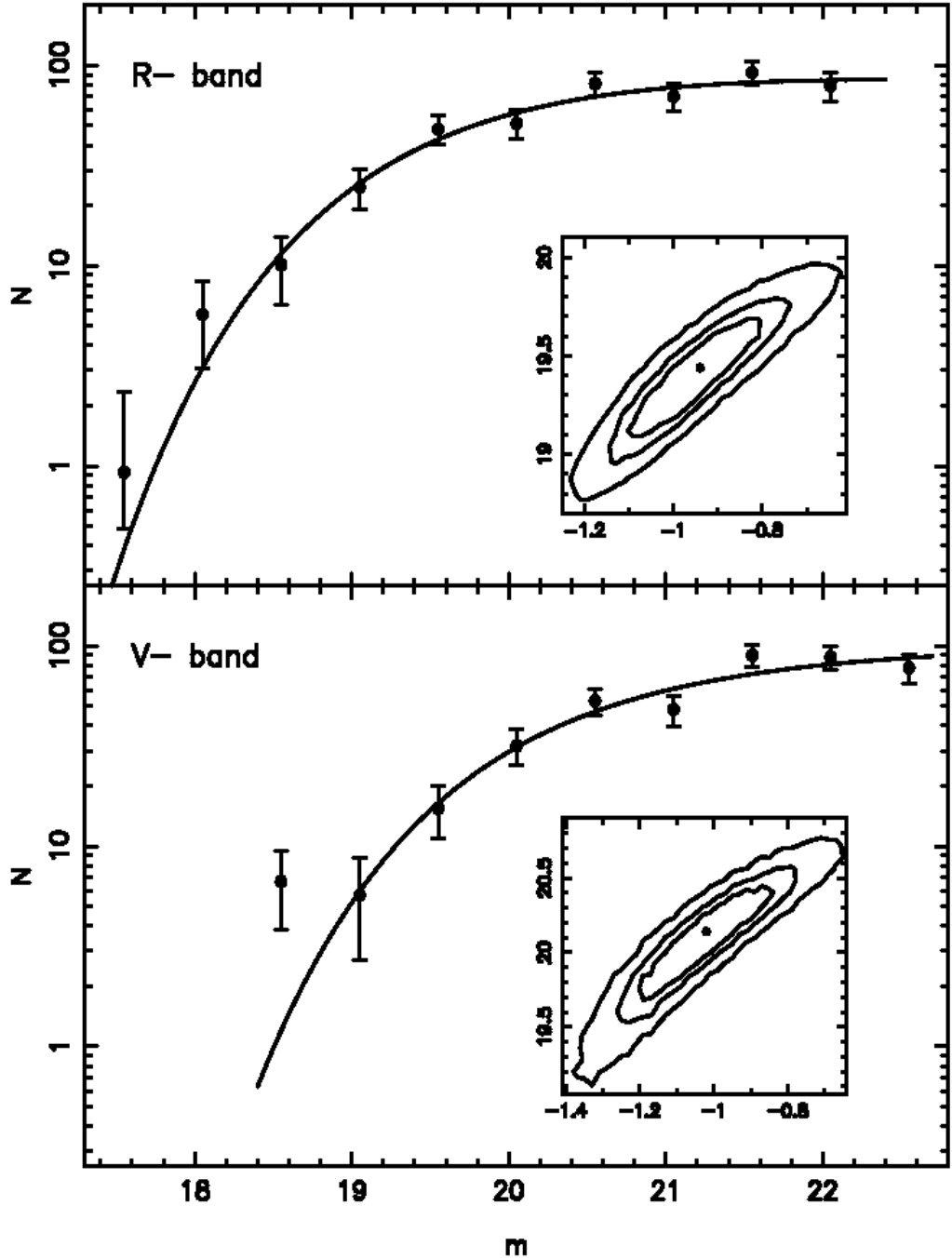}}
\caption{Luminosity functions in the R- and V-bands. In the smaller
panels the confidence levels of the best--fit parameters $\alpha$ (in
abscissa) and $M^*$ are shown. The contour values correspond to 
the confidence levels of 68\%, 90\% and 99\% for a normal random deviate.}
\label{LF_FIG}
\end{figure}

The errors on the cluster LFs were computed by taking into account Poissonian
fluctuations of counts along the cluster line of sight and of the
background counts.  Since the field of view of
our observation is $\sim 15$ times smaller than the area covered by
the ESO--Sculptor Survey, the errors on the cluster LFs are dominated
by Poissonian errors of counts along the cluster line of sight.  
For this reason we did not include
in the error budget the term of the background fluctuations due to
the galaxy correlation function.

We modeled the cluster LFs with a weighted parametric fit of the
Schechter function, obtaining as best--fit parameters: 
$R^*=19.4\pm 0.3$ and $\alpha_R=-0.94\pm0.16$, 
$V^*=20.1\pm0.4$ and $\alpha_V=-1.02\pm0.18$ 
(the errors correspond to 1$\sigma$ interval). 
Fig.~\ref{LF_FIG} shows the cluster LFs
respectively in R-band (upper panel) and V-band (lower panel), with
the confidence levels of the best--fit parameters. The best--fitting
Schechter functions are also plotted.

Andreon (2001) derived the K-band LF in AC\,118. On the whole K-band
field of $\sim 6^\prime \times 6^\prime$, he found: $K^*=15.3$ and
$\alpha_K=-1.2$. Taking into account the uncertainties on the $\alpha$
parameter (see our Fig.~\ref{LF_FIG} and Fig. 9 of Andreon 2001), we
conclude that $\alpha_R$, $\alpha_V$ and $\alpha_K$ are in
agreement (within the errors), that is the faint end slope of the
AC\,118 LF does not depend significantly on the waveband. 
Choosing $\alpha_R=\alpha_V=\alpha_K=-1.1$ we obtain $(V^*-K^*)=4.2
\pm 0.3$ and $(R^*-K^*)=3.4 \pm 0.3$, in agreement with the position
of the bright tail of the galaxy population in the relative
colour--magnitude planes (Merluzzi et al. 2002; Merluzzi et al. 2002,
in preparation).

\section{Summary}
New accurate $\mathrm{VRI}$ photometry has been presented for the
galaxy cluster AC\,118 at redshift $z=0.31$. The new data
cover an area of 8.6$\times$8.6 arcmin$^2$, corresponding to
2.9$\times$2.9 Mpc$^2$.
The data reduction and the method used for the photometric calibration
are described in detail. For the R-band, the image zero-point
is derived by using aperture photometry by SED98, while for the V- and 
I-band the photometric
calibration is performed by matching the colors of the stars in the
cluster field with those of the stars from the catalogue by Landolt
(1992). The accuracies on the zero-points amount to $\sim0.04$ mag. 

For each band, we derive magnitudes within a fixed aperture of
diameter 4.4$''$ and Kron (adaptive) magnitudes. 
The catalogue is complete up to $\mathrm{V}=22.8$ mag,
$\mathrm{R}=22.3$ mag and $\mathrm{I}=20.8$ mag, and
includes, respectively, $\mathrm{N}=$ 574, 649 and 419 galaxies 
brighter than these limits.

The cross-correlated catalogue includes celestial coordinates, fixed
aperture and Kron magnitudes, and Kron radii of $\mathrm{N}=1206$
sources detected in at least one band. The catalogue is available in
electronic form.  For the $\mathrm{N}=459$ sources with additional U-
and K-band photometry, we also obtain a redshift estimate by applying
the photometric redshift technique.

The V- and R-band photometry have been
used to derive the optical luminosity function of the cluster AC\,118.
The best-fit Schechter parameters are $R^*=19.4\pm0.3$ and 
$\alpha_R=-0.94\pm0.16$,
$V^*=20.1\pm0.4$ and $\alpha_V=-1.02\pm0.18$. Comparing these values
with the results obtained by Andreon (2001) for the K-band LF of
AC\,118, we conclude that the faint end slope of the cluster LF does
not depend on the waveband, while the waveband dependence of $M^{*}$
is consistent with pure passive evolution of the stellar populations
of the bright cluster galaxies.

\begin{acknowledgements}
We are grateful to Adam Stanford who provided us with his own photometry 
of AC\,118, that was used for the calibration of our R-band data
and to Stephane Arnouts who provided us with the galaxy counts used to
derive the cluster LFs. We thank S. Zaggia for helpful
discussions. We thank the referee A. Mazure for his helpful comments.
The observations at ESO were collected during the NTT guaranteed time of 
the Osservatorio Astronomico di Capodimonte. The U-band photometry is based
on data from the ESO NTT Science Archive.
Michele Massarotti is partly supported by a MURST-COFIN grant.
\end{acknowledgements}

\end{document}